\documentclass[10pt,twocolumn,showpacs,preprintnumbers,amsmath,amssymb,aps,prl,superscriptaddress,longbibliography]{revtex4-2}
\usepackage{appendix}
\usepackage{bbm}
\usepackage{mathrsfs}
\usepackage{graphicx}
\usepackage{dcolumn}
\usepackage{bm}
\usepackage{braket}
\usepackage{amsmath}
\usepackage{amsfonts}
\usepackage[dvipsnames]{xcolor}
\usepackage{tabularray}
\usepackage[colorlinks=true,linkcolor=Blue,urlcolor=BlueViolet,citecolor=BlueViolet]{hyperref}

\begin{document}

\title{Vortex-Number-Controlled Josephson Diode Polarity in Corbino Junctions}
\author{Linghao Huang}
\affiliation{State Key Laboratory of Surface Physics and Department of Physics, Fudan University, Shanghai 200433, China} 
\affiliation{Shanghai Research Center for Quantum Sciences, Shanghai 201315, China}
\author{Jing Wang}
\thanks{wjingphys@fudan.edu.cn}
\affiliation{State Key Laboratory of Surface Physics and Department of Physics, Fudan University, Shanghai 200433, China}
\affiliation{Shanghai Research Center for Quantum Sciences, Shanghai 201315, China}
\affiliation{Institute for Nanoelectronic Devices and Quantum Computing, Fudan University, Shanghai 200433, China}
\affiliation{Hefei National Laboratory, Hefei 230088, China}

\begin{abstract}
We demonstrate that the polarity of the Josephson diode effect in Corbino Josephson junctions can be deterministically controlled by the Josephson-vortex number, which arises as a generic consequence of structured spatial inhomogeneity. By solving the continuum Andreev spectral problem, we identify a mechanism in which the vortex number selectively filters specific spatial Fourier harmonics of the local inhomogeneity. This harmonic selection reshapes the amplitudes and relative phases of higher-order Josephson current harmonics, ultimately reversing the critical-current asymmetry. Numerical simulations of both an effective one-dimensional edge model and full two-dimensional lattice models confirm the robustness of this mechanism. Crucially, our results show that a vortex-parity-dependent reversal of diode polarity is not an exclusive signature of Majorana physics, but can emerge generically from geometric and structural inhomogeneities in Corbino junctions.
\end{abstract}

\maketitle

\emph{Introduction}---The superconducting diode effect~\cite{ando2020observation,miyasaka2021observation,daido2022intrinsic,yuan2022supercurrent,he2022phenomenological,narita2022fieldfree,tanaka2022theory,jiang2022fieldfree,nadeem2023superconducting,hou2023ubiquitous,depicoli2023superconducting}, a nonreciprocal superconducting transport response featuring asymmetric forward ($I_c^+$) and backward ($I_c^-$) critical currents, has emerged as a powerful probe of symmetry breaking in quantum materials. A prominent realization is the Josephson diode effect (JDE)~\cite{hu2007proposed,misaki2021theory,davydova2022universal,zhang2022general,baumgartner2022supercurrent,wu2022fieldfree,pal2022josephson,liu2013manipulating,nagaosa2024nonreciprocal,
aligia2020tomography,wu2022fieldfree,kopasov2021geometry,ilic2022theory,diez-merida2023symmetrybroken,halterman2022supercurrent,karabassov2022hybrid,gupta2023gatetunable,turini2022josephson,kokkeler2022fieldfree,lu2023tunable,mazur2024gatetunable,hu2023josephson,fu2024fieldeffect,trahms2023diode,steiner2023diode,banerjee2023phase,legg2023parityprotected,maiani2023nonsinusoidal,cheng2023josephson,lotfizadeh2024superconducting,chen2024edelstein,costa2023microscopic,cuozzo2024microwavetunable,ciaccia2023gatetunable,matsuo2023josephson,valentini2024parityconserving,pillet2023josephson,seoanesouto2024tuning,debnath2024gatetunable,fukaya2025supercurrent,yerin2024supercurrent,scharf2024superconducting,shen2025josephson,kotetes2026nonreciprocal,costa2025unconventional,nikodem2025tunable,yerin2025supercurrent,kudriashov2025nonmajorana} in Josephson junctions~\cite{likharev1979superconducting}, where the superconducting phase difference offers an additional knob for engineering nonreciprocal supercurrents. The JDE 
fundamentally requires the lifting of certain underlying symmetries~\cite{zinkl2022symmetry,wang2025currentreversion}, which skews the phase-dependent Andreev spectrum. This spectral distortion induces distinct phase shifts among different harmonics of the current-phase relation (CPR)~\cite{baumgartner2022supercurrent,souto2022josephson,fominov2022asymmetric,pal2022josephson}, ultimately driving a nonreciprocal critical current.

\begin{figure}[t]
  \begin{center}
  \includegraphics[width=3.4in,clip=true]{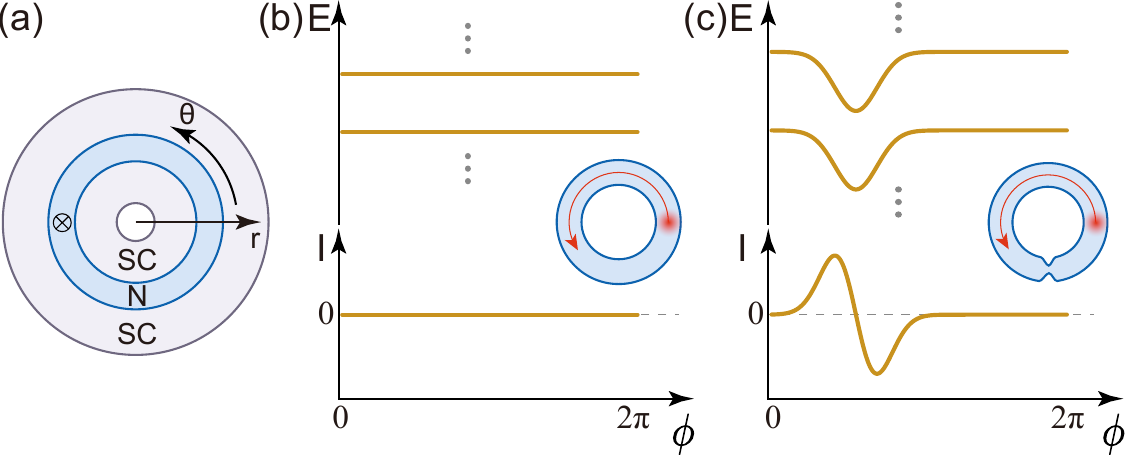}
  \end{center}
  \caption{Corbino Josephson junction and current-phase relation (CPR). (a) Schematic of a Corbino Josephson junction. A perpendicular magnetic field threads the non-superconducting region (N), enclosing a magnetic flux $\Phi$ quantized in units of the superconducting flux quantum $\Phi_0=h/2e$. (b) Andreev spectrum and CPR of a homogeneous junction with $\Phi=\Phi_0$. (c) Corresponding results for an inhomogeneous junction. The diffuse red spot denotes a Josephson vortex, which moves around the annulus as the phase difference $\phi$ between the inner and outer superconductors is increased.}
  \label{fig1}
\end{figure}

While most investigations have focused on linear or planar junctions, Corbino Josephson junctions~\cite{hadfield2003corbino,clem2010corbinogeometry,park2015detecting,park2020electrontunnelingassisted,matsuo2020evaluation,zhang2022ac,okugawa2022vortex,piasotski2025current,lesser2026theory,rycerz2026josephson} offer a complementary closed-loop geometry [Fig.~\ref{fig1}(a)] where supercurrent flows radially between concentric inner and outer superconducting contacts through an annular weak link. This boundaryless configuration eliminates contributions from sample-edge current paths and enforces a strict topological constraint via fluxoid quantization~\cite{tinkham2004introduction}. An out-of-plane magnetic field establishes an integer winding number $p=\Phi/\Phi_0$, dictating the number of Josephson vortices trapped within the annulus. Because topological Josephson junctions can harbor unconventional transport signatures distinct from their trivial counterparts~\cite{fu2008superconducting,lutchyn2010majorana,oreg2010helical,alicea2010majorana,hell2017twodimensional,pientka2017topological}, nonreciprocal phenomena in topological junctions have recently attracted intense attention~\cite{cayao2024enhancing,liu2024josephson,mondal2025josephson,wang2026giant,sten2026josephson,hou2026gatetunable}. In particular, a striking experiment on Corbino junctions fabricated on three-dimensional topological insulators revealed that the diode polarity alternates deterministically with the parity of the enclosed vortex number~\cite{park2026vortexparitycontrolled}:
\begin{equation}
    \operatorname{sgn}(\eta)=(-1)^{p},
\end{equation}
where $\eta=\frac{I_c^+ - |I_c^-|}{I_c^+ + |I_c^-|}$ is the diode efficiency factor. This behavior was originally attributed to a purely topological origin: a vortex-parity-induced switch between periodic and antiperiodic boundary conditions for Majorana edge states. However, because the JDE is fundamentally born out of symmetry breaking rather than topology, this observation poses a critical question: is vortex-parity-dependent polarity a unique smoking gun of Majorana physics, or can it be emulated by trivial, ubiquitous mechanisms?

In this Letter, we demonstrate that structured spatial inhomogeneity provides a universal, non-topological route to such polarity alternation. In an ideal, rotationally symmetric Corbino junction [Fig.~\ref{fig1}(b)], shifting the global phase difference $\phi$ between the outer and inner superconductors simply rotates the vortex texture around the annulus, leaving the Andreev levels completely invariant and the net Josephson current zero. Spatial inhomogeneity breaks this rotational equivalence; consequently, the vortex texture samples a position-dependent potential landscape, and its evolution around the annulus modulates the Andreev spectrum to yield a finite, phase-dependent CPR [Fig.~\ref{fig1}(c)]. By decomposing this angular inhomogeneity into Fourier components, we establish a harmonic-selection rule: the vortex number dictates which spatial harmonics map into phase harmonics of the Andreev spectrum. These selected spectral harmonics generate vortex-number-dependent current harmonics, ultimately reversing the critical-current asymmetry when their relative phase shifts are appropriately arranged. We verify the robustness of this mechanism in both topological and non-topological Corbino junctions. Our results recast vortex-parity-dependent Josephson diode responses as a consequence of vortex-number harmonic selection within the Andreev spectrum, offering a powerful strategy to design quantum nonreciprocity through structured inhomogeneity.

\emph{Andreev spectral analysis}---We begin with a low-energy description of the Josephson junction. A minimal two-component Hamiltonian reads $\mathcal{H}_\text{eff}(\phi)=\int_0^{2\pi} \mathrm{d}\theta\, H(\theta,\phi)$, \begin{equation}\label{ana_H}
  H(\theta,\phi) = i v \frac{\partial}{\partial \theta}\sigma_z - \Delta(\theta) \cos\left(\frac{\phi+p\theta}{2}\right) \sigma_y,
\end{equation}
where $\theta$ is the polar angle, $v$ is the velocity, and $p=\Phi/\Phi_0$ is the vortex number. The combination $\phi+p\theta$ is the gauge-invariant phase difference in a gauge where the enclosed flux appears as an angular winding~\cite{tinkham2004introduction,potter2013anomalous}. The Pauli matrices $\sigma_{y},\sigma_z$ act in the two-component basis. The first kinetic term describes two counterpropagating modes, while the second is the phase-dependent Andreev coupling. This formulation applies equally to topological and non-topological junctions. In the former, the basis $[\gamma_1(\theta),~\gamma_2(\theta)]^\text{T}$ represents Majorana fields~\cite{fu2008superconducting}, whereas in the latter it describes a perfectly transmitting Andreev channel in the pseudo-Nambu basis $[\Psi_{\mathrm{in},\uparrow}(\theta),\Psi_{\mathrm{out},\downarrow}^{\dagger}(\theta)]^\text{T}$, and the two components represent the incoming electron and outgoing hole channels coupled by Andreev reflection~\cite{ivanov1999twolevel}. The pairing profile is periodic along the annulus and can be expanded as
\begin{equation}\label{Delta}
  \Delta(\theta)\equiv\Delta_0+\delta\Delta(\theta)=\Delta_0+\sum\limits_{j>0} \Delta_j \cos(j\theta+\varphi_j),
\end{equation}
where $\Delta_0,\Delta_j>0$, $0\le\varphi_j<2\pi$. Here we consider a weak spatial modulation $\Delta_j\ll\Delta_0$. At zero temperature, the equilibrium Josephson current is given by~\cite{beenakker1991josephson,golubov2004currentphase}
\begin{equation}\label{current}
  I(\phi)=-\frac{2e}{\hbar}\sum_{E_n>0} \frac{\partial E_n(\phi)}{\partial \phi},
\end{equation}
where $E_n(\phi)$ satisfies $H(\theta,\phi)\Psi_n(\theta,\phi)=E_n(\phi)\Psi_n(\theta,\phi)$. 

A key consequence of the Corbino geometry is a harmonic-selection rule: only spatial Fourier components $j=\ell p$ of $\delta\Delta(\theta)$ generate the $\ell$-th harmonic of the current-phase relation, while mismatched components do not contribute to the phase-dependent spectrum at leading order. To make this explicit, we combine symmetry analysis with first-order perturbation theory. For a homogeneous annulus, a change in $\phi$ is equivalent to a translation of the Hamiltonian $H_0(\theta,\phi)=\mathcal T_{\phi/p}^{-1}H_0(\theta,0)\mathcal T_{\phi/p}$, with $\mathcal T_a\Psi(\theta,\phi)\equiv\Psi(\theta-a,\phi)$, implying that the unperturbed spectrum $E_n^{(0)}$ is $\phi$-independent. Introducing the co-moving coordinate $x=\theta+\phi/p$, a weak modulation produces the first-order shift in energy
\begin{equation}\label{perturb}
  \delta E_n^{(1)}(\phi)=\int_{0}^{2\pi} \mathrm{d}x ~ W_n(x) \delta\Delta\left(x-\frac{\phi}{p}\right),
\end{equation}
where $W_n(x)=-\cos(px/2)\Psi_n^\dagger(x)\sigma_y\Psi_n(x)$ is the modified density profile of the unperturbed state. This profile has a $p$-fold rotational symmetry: under $x\to x+2\pi/p$, both $\cos(px/2)$ and $\Psi_n^\dagger\sigma_y\Psi_n$ change sign, leaving $W_n(x)$ invariant~\cite{sm}. Consequently, its Fourier expansion contains only harmonics of wave number $\ell p$:
\begin{equation}
  W_n(x)=W_{n,0}+\sum_{\ell>0}W_{n,\ell}\cos(\ell p x+\alpha_{n,\ell}),
\end{equation}
where $W_{n,\ell}>0$ is the amplitude and $\alpha_{n,\ell}$ is the corresponding phase. Expanding $\delta\Delta(\theta=x-\phi/p)$  in Fourier harmonics and performing the angular integration in Eq.~(\ref{perturb}) yields a strict matching condition between Fourier components of $W_n(x)$ and $\delta\Delta(\theta)$:
\begin{equation}
\delta E_n^{(1)}(\phi)=\pi\sum_{\ell>0}W_{n,\ell}\Delta_{\ell p}\cos\left(\ell\phi+\alpha_{n,\ell}-\varphi_{\ell p}\right)+\mathrm{const.}
\end{equation}
The corresponding current contribution follows as
\begin{equation}
I_{n,\ell}^{(1)}(\phi) \propto\ell W_{n,\ell}\Delta_{\ell p}\sin\left(\ell\phi+\alpha_{n,\ell}-\varphi_{\ell p}\right),
\end{equation}
demonstrating explicitly that only $j=\ell p$ components generate $\phi$-dependent energy corrections and produce the $\ell$-th Josephson current harmonic.

In the local bound-state limit, this selection rule can be verified directly from the analytic solution near zeros of the Andreev coupling (see End Matter). Beyond the selection rule, the relative strength of the allowed harmonics is set by the Fourier content of $W_n(x)$. For higher-energy quasi-continuum states, the kinetic term dominates, and $\Psi_n(x)$ becomes progressively delocalized along the annulus. This smoothens $W_n(x)$, suppressing higher Fourier components and thereby reducing higher-order current harmonics.

\begin{figure}[t]
  \begin{center}
  \includegraphics[width=3.4in,clip=true]{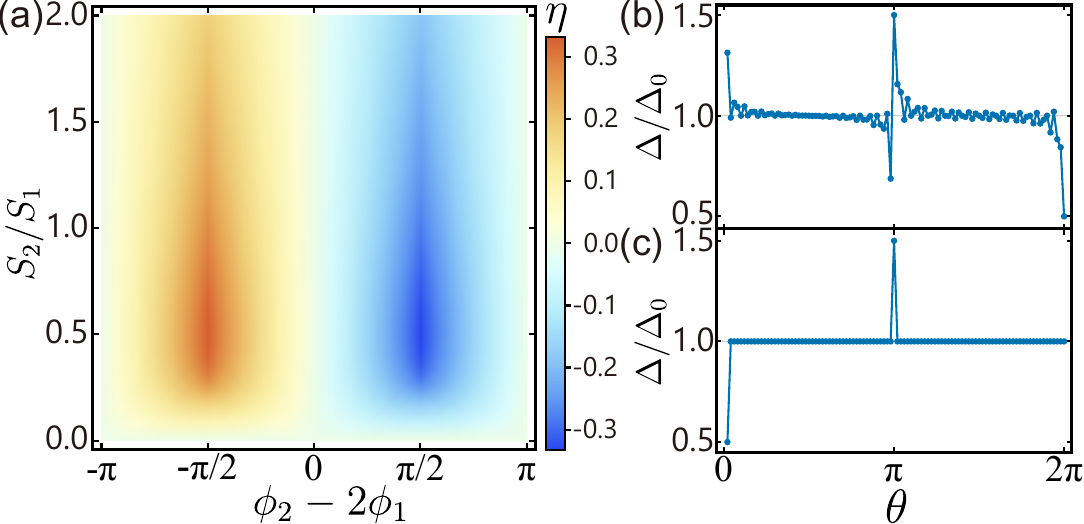}
  \end{center}
  \caption{The Josephson diode polarity and representative inhomogeneous pairing profile $\Delta(\theta)$. (a) Diode polarity obtained from Eq.~(\ref{1st2nd}). (b),(c) Two representative examples of $\Delta(\theta)$ evaluated on a uniform discretization of $\theta \in [0,2\pi)$ with $100$ grid points. Panel (b) corresponds to the profile defined in Eq.~(\ref{Delta}) with $n_{\mathrm{c}}=60$. (c) shows a two-site perturbation profile, in which the pairing amplitude exhibits opposite-sign deviations from the uniform background at two nearly diametrically separated grid points (sites $1$ and $50$).}
  \label{fig2}
\end{figure}

\emph{Diode polarity}---We now analyze the origin of alternating diode polarity. Since higher harmonics are suppressed, we retain the lowest two nonzero contributions to the current:
\begin{equation}\label{1st2nd}
  I(\phi)=S_1\sin(\phi+\phi_1)+S_2\sin(2\phi+\phi_2),
\end{equation}
where $S_1,S_2>0$, and $\phi_1,\phi_2$ are the phase shifts of the first and second current harmonics. The diode polarity is controlled by their relative phase. A direct evaluation (see End Matter) yields
\begin{equation}\label{sgneta}
\operatorname{sgn}(\eta)=-\operatorname{sgn}\left[\sin(\phi_2-2\phi_1)\right].
\end{equation}
Hence, alternating diode polarity requires $\phi_2-2\phi_1$ to alternate with the vortex number $p$. Fig.~\ref{fig2}(a) shows $\eta$ obtained from Eq.~(\ref{1st2nd}), confirming maximum alternation for $\phi_2-2\phi_1=\pm\pi/2$. The relative CPR phases are directly determined by the Fourier phases of the pairing profile~\cite{sm}: $-\operatorname{sgn}\left[\sin(\phi_2-2\phi_1)\right] =\xi_p\operatorname{sgn}\left[\sin(\varphi_{2p}-2\varphi_p)\right]$, with $\xi_p=\pm1$ typically retaining the same sign across $p$. Thus, generically, $\operatorname{sgn}(\eta)=\xi\operatorname{sgn}\left[\sin(\varphi_{2p}-2\varphi_p)\right]$, where $\xi=\pm1$ is a model-dependent but $p$-independent sign factor. Thus, achieving $\varphi_{2p}-2\varphi_p=(-1)^p(\pi/2)$ yields the maximal alternation of $\eta$. We therefore construct a representative inhomogeneity profile $\delta\Delta(\theta)$ designed to produce a vortex-number-controlled alternating sign of $\eta$. Starting from Eq.~(\ref{Delta}), we choose $\varphi_j=0$ for odd $j$ and $\varphi_j=\pi/2$ for even $j$, with all $\Delta_j$ taken equal. Introducing a cutoff $j\le n_{\mathrm{c}}$, the series can be summed analytically, giving
\begin{equation}\label{harmonic}
  \delta\Delta(\theta) = \frac{\Delta_0\sin(n_\text{c}\theta)}{2n_\text{c}\sin(\theta)}\{\cos(n_\text{c}\theta)+\sin[(n_\text{c}+1)\theta]\}.
\end{equation}
Fig.~\ref{fig2}(b) shows this profile sampled on a uniform angular grid, corresponding to the discrete values $\Delta_{n_i}$ used in the numerics. Its dominant feature is two opposite-sign modulations localized in two nearly diametrically opposite angular sectors. Motivated by this structure, we further consider a simplified two-site perturbation profile with opposite-sign pairing-amplitude perturbations at two nearly diametrically separated grid sites on the ring [Fig.~\ref{fig2}(c)]. This reduced profile introduces only small deviations in $\varphi_j$, which weakly renormalize the harmonic phases but do not affect the vortex-number-controlled alternation of the diode polarity.

\begin{figure}[b]
  \begin{center}
  \includegraphics[width=3.4in,clip=true]{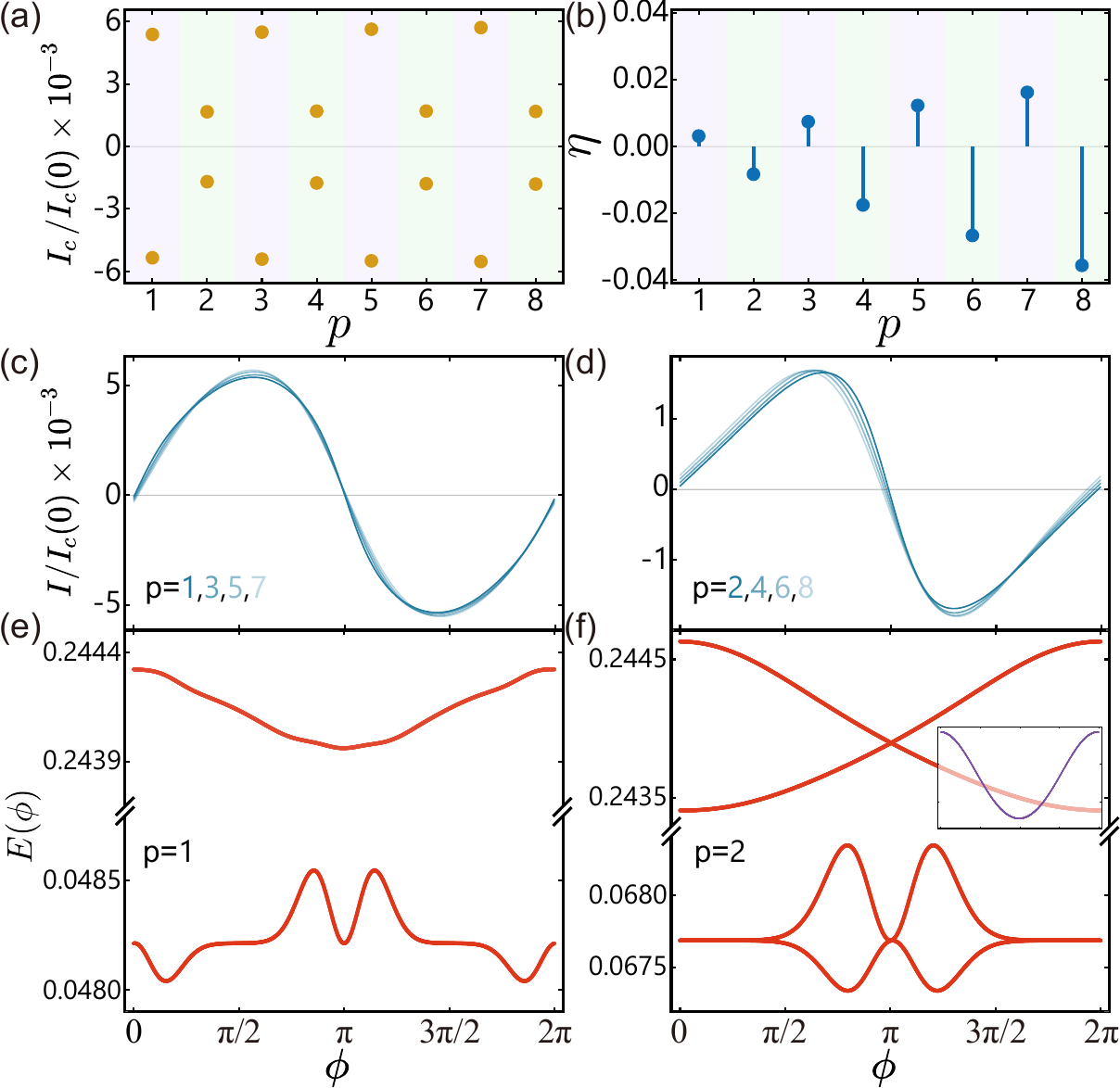}
  \end{center}
  \caption{Numerical results for effective 1D edge model. (a) Forward and backward critical currents $I_c^\pm$ as a function of vortex number $p$. (b) Corresponding diode polarity $\eta$ versus $p$. (c),(d) Current-phase relations for odd and even $p$, respectively. (e),(f) Representative low-energy and quasi-continuum Andreev spectra for $p=1$ and $p=2$, respectively. The color scale indicates the wavefunction weight on the junction Majorana modes. The inset in (f) shows the sum of the two quasi-continuum branches. Model parameters are $v=1,~\Delta_0=0.15,~m=0.6,~E_\text{cut}=0.5$.}
  \label{fig3}
\end{figure}

\emph{Numerical simulation: 1D edge-model---}To substantiate the above physical picture, we next consider a minimal model in which the full spectrum can be computed explicitly. We employ a Majorana edge model describing a topological Josephson junction, where two counterpropagating Majorana modes reside at the two superconductor--normal (S--N) interfaces. Because the Nielsen-Ninomiya theorem forbids an isolated chiral Majorana mode on a one-dimensional lattice~\cite{nielsen1981nogo}, we adopt a variant of the Grover-Sheng-Vishwanath construction~\cite{grover2014emergent}. Specifically, one pair of Majorana modes is introduced on each edge, and a Wilson mass $m$ gaps the crossing at the Brillouin-zone boundary. The two counterpropagating modes at the junction are coupled by the Josephson term, while the remaining vacuum-boundary modes stay gapless. The uniform zero-flux Hamiltonian is defined by
\begin{equation}
  \begin{aligned}
    H^{\text{edge}}_0(k)=&\tau_0\otimes\left[v\sin k~\sigma_z+m(1-\cos k)\sigma_y\right] \\
    &+\frac{\Delta_0}{2}\cos\left(\frac{\phi}{2}\right)\left(\tau_x\otimes\sigma_y-\tau_y\otimes\sigma_x\right),
  \end{aligned}
\end{equation}
where $k$ is the angular wavenumber and $v$ is the velocity of Majorana states. For the flux-threaded inhomogeneous junction, we use the corresponding real-space lattice Hamiltonian and replace the uniform Josephson coupling by the site-dependent term $\Delta_{n_i}\cos[(\phi+p\theta_{n_i})/2]$, as dictated by Eq.~(\ref{ana_H}). We impose periodic boundary conditions on all states, except that the two outer-edge Majorana states are antiperiodic for odd $p$. We adopt the two-site texture of $\Delta_{n_i}$ discussed above by discretizing the annulus into 400 sites and choosing $\Delta_{1}=2\Delta_0$ and $\Delta_{200}=0.5\Delta_0$. To evaluate $I(\phi)$ via Eq.~(\ref{current}), we remove the spurious contribution of the vacuum-boundary Majorana modes by restricting the spectral sum to $E_n<E_\text{cut}$~\cite{lesser2026theory}.

Figures~\ref{fig3}(a) and~\ref{fig3}(b) show the critical currents $I_c^\pm$ and diode polarity $\eta$ for vortex numbers up to $p=8$, while Figs.~\ref{fig3}(c) and~\ref{fig3}(d) display the corresponding CPRs. The results exhibit a clear vortex-parity effect: the sign of $\eta$ alternates with $p$. Moreover, the CPRs collapse into two distinct families according to the parity of $p$, consistent with the alternating diode polarity.

Representative Andreev spectra are shown in Figs.~\ref{fig3}(e) and~\ref{fig3}(f) for $p=1$ and $p=2$, respectively. For $p=1$, the low-energy bound states sample the two localized pairing modulations as they circulate around the annulus, leading to pronounced spectral modulation near $\phi=0$ and $\pi$~\footnote{We choose the first positive Andreev level in Fig.~\ref{fig3}(e) and~\ref{fig3}(f), so the detailed modulation of the spectrum near $\phi=0$ and $\pi$ resembles the density distribution of the first excitation state, as can be understood via first-order perturbation analysis.}. The quasi-continuum states exhibit the expected cosine-like phase dependence. For $p=2$, the two nearly degenerate bound states translate by only $\pi$ during one $2\pi$ phase cycle of $\phi$ and therefore probe complementary halves of the pairing profile. Consequently, one branch disperses upward while the other disperses downward. The quasi-continuum states display an analogous structure: two complementary branches phase-shifted from each other by approximately $\pi$, each tracing roughly one half-cycle of a cosine-like modulation. In analogy with the treatment of local degenerate states in End Matter, summing each nearly degenerate pair restores an approximately cosine-like dependence, as shown in the inset of Fig.~\ref{fig3}(f). These spectral features provide direct numerical evidence for the vortex-number harmonic-selection mechanism discussed above.

\emph{Numerical simulation: 2D model---}We next test the proposed mechanism in full two-dimensional lattice models. This provides a microscopic validation and enables a direct comparison between topological and non-topological junctions on equal footing. We consider a rectangular superconductor--normal--superconductor (SNS) geometry, with width $W$ along the $y$ direction. The length of the N region is $L$. The two S regions are taken to be semi-infinite to suppress finite-size effects along the $x$ direction. We discretize the structure into a square lattice with unit lattice constant, and impose periodic boundary conditions along the $y$ direction. This setup corresponds to a large-radius Corbino junction. The equilibrium Josephson current is computed from the bond-current expectation value across the junction~\cite{sm}.

We use the following uniform, zero-flux Bloch Hamiltonians to define the microscopic models, while the simulations are performed in real space with Peierls phases and a position-dependent pairing function. For the homogeneous S region, the Hamiltonian written in the Bogoliubov-de Gennes (BdG) form is
\begin{equation}\label{BdG2D}
  H^{\text{2D}}(\mathbf{k}) =
    \begin{pmatrix}
      H_0(\mathbf{k}) & \Delta(\mathbf{k})\\
      \Delta(\mathbf{k})^\dag & -H_0^*(-{\mathbf{k}})
    \end{pmatrix},
\end{equation}
with pairing removed in the N region. For the topological case, we model superconducting proximity on the surface of a ferromagnetic topological insulator~\cite{wang2015chiral,he2019platform}. The normal-state Hamiltonian is $H_0(\mathbf{k})=v(k_y\sigma_x\tau_z-k_x\sigma_y\tau_z)+M(\mathbf{k})\tau_x+m_z\sigma_z-\mu$ in the basis $(c^t_{\mathbf{k}\uparrow}, c^t_{\mathbf{k}\downarrow},c^b_{\mathbf{k}\uparrow}, c^b_{\mathbf{k}\downarrow})^\text{T}$. Here the superscripts $t$ and $b$ denote the top and bottom surface states, respectively. $\sigma_i$ and $\tau_i$ are Pauli matrices for spin and layer, respectively. $m_z$ is the Zeeman field. $M(\mathbf{k})=M_0+M_1(k_x^2+k_y^2)$ describes inter-surface hybridization. $\mu$ is the chemical potential. The induced pairing $\Delta(\mathbf{k}) = i \Delta_0 \sigma_y(\tau_0+\tau_z)/2$ acts only on the top surface layer. In the parameter regime corresponding to BdG Chern number $\mathcal{C}=1$, the model hosts a pair of counterpropagating Majorana edge states localized at the two S--N interfaces.

For comparison, the non-topological case is described by a single-band square-lattice model  $H_0(\mathbf{k})=2t(2-\cos k_x-\cos k_y) - \mu$ with conventional $s$-wave pairing. In both cases, spatial inhomogeneity is introduced by modulating the pairing amplitude along $y$ using the Fourier-engineered profile $\delta\Delta$ defined in Eq.~(\ref{harmonic}).

\begin{figure}[t]
  \begin{center}
  \includegraphics[width=3.4in,clip=true]{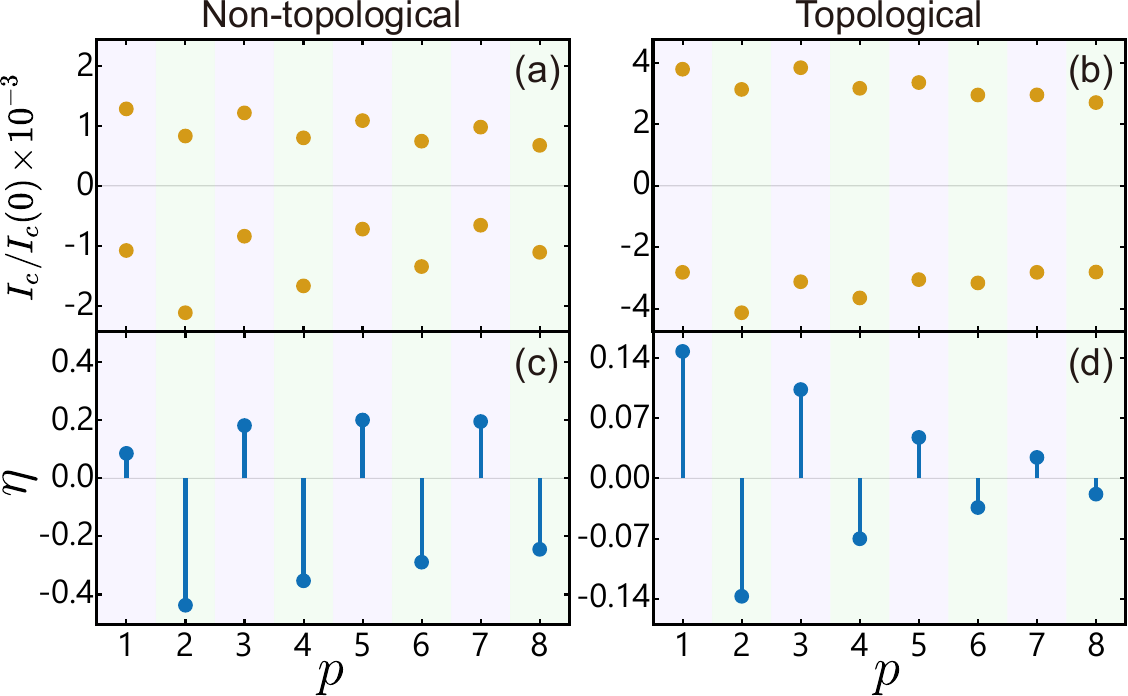}
  \end{center}
  \caption{Numerical results for 2D models. Panels (a),(c) show the forward and backward critical currents $I_c^\pm$ and diode polarity $\eta$, respectively, for a non-topological Corbino Josephson junction. The model parameters are $W=400,~L=2,~t=20,~\mu=10,~\Delta_0=1,~n_\text{c}=300$. Panels (b),(d) show the corresponding results for a topological junction. The model parameters are $W=300,~L=2,~t=300,~M_0=20,~M_1=300,~m_z=20,~\mu=5,~\Delta_0=10,~n_\text{c}=200$.}
  \label{fig4}
\end{figure}

Figure~\ref{fig4} shows the resulting critical currents $I_c^\pm$ and diode polarity $\eta$. Both the topological and non-topological models exhibit clear alternation of $\mathrm{sgn}(\eta)$ with vortex number, demonstrating that the effect is controlled by the structured pairing profile rather than the presence of Majorana modes. The behavior is robust provided that the relative phases of the selected Fourier components satisfy the alternation condition discussed above. Additional numerical results are provided in the Supplemental Material.

\emph{Discussion---}While our continuum model in Eq.~(\ref{ana_H}) describes a non-topological SNS junction in the perfect-transmission limit, realistic devices inevitably feature finite interface transparency due to material mismatches and fabrication-induced disorder. Crucially, finite transparency primarily renormalizes the amplitudes of the CPR harmonics and suppresses higher-order harmonics, but it does not alter the symmetry constraint underlying the harmonic-selection rule. We confirm the robustness of this mechanism against reduced transparency via numerical simulations in Supplemental Material~\cite{sm}.

The mechanism proposed here relies on two essential ingredients: the Corbino geometry and a structured inhomogeneity. The closed-loop geometry eliminates sample-edge current paths and promotes the vortex number to a discrete control parameter, while the structured inhomogeneity provides spatial Fourier components whose phases are inherited by the selected CPR harmonics. Their interplay enables deterministic control of the diode polarity by the vortex number. Importantly, this mechanism does not require a topological origin, as demonstrated by our simulations of non-topological junctions. At the same time, our results do not exclude additional topology-specific mechanisms that may contribute to alternating diode polarity in topological junctions.

Although we focused on a spatial modulation of the pairing profile as a representative form of structured inhomogeneity, our selection rule is generic. Variations in junction transparency, local chemical potential, or weak-link width can similarly induce harmonic selection and vortex-number-controlled diode responses, despite differing in their microscopic origin. Experimentally, such structures may be engineered by local electrostatic gates that tune the carrier density or barrier height, patterned superconducting coverage that modifies the proximity strength, or lithographically defined variations of the annular weak-link width~\cite{heersche2007bipolar,elfeky2021local}. Motivated by the recent observation of vortex-parity-controlled diode polarity in Corbino junctions~\cite{park2026vortexparitycontrolled}, a direct experimental test of our mechanism could be realized by placing two local electrostatic gates at nearly diametrically opposite positions on the annulus. Applying opposite gate voltages would realize an effective two-site perturbation profile, while reversing the gate polarity should directly invert the vortex-parity pattern of the diode signal.

Finally, structured inhomogeneity can be exploited to engineer complex, vortex-number-controlled diode responses well beyond the simple binary parity pattern $\operatorname{sgn}(\eta)=(-1)^p$. The key design principle is to control the selected Fourier-phase combination $\varphi_{2p}-2\varphi_p$, which uniquely dictates the sign of the diode polarity. For instance, a four-vortex-periodic pattern,
\begin{equation}
  \operatorname{sgn}(\eta)=(-1)^{\lfloor (p-1)/2\rfloor},
\end{equation}
with $\lfloor x\rfloor$ denotes the greatest integer not exceeding $x$, can be targeted by tailoring the spatial inhomogeneity such that $\varphi_{2p}-2\varphi_p=\frac{\pi}{2}(-1)^{\lfloor (p-1)/2\rfloor}$. This can be explicitly realized via the harmonic phase configuration $\varphi_{2n-1}=0$, $\varphi_{2n}=\frac{\pi}{2}(-1)^{\lfloor n/2\rfloor}$ (with $n\in\mathbb{N}^+$). Structured inhomogeneity therefore provides a customizable route to programmable vortex-number-dependent Josephson nonreciprocity.

\begin{acknowledgments}
\emph{Acknowledgments}---This work is supported by the National Key Research Program of China under Grant No.~2025YFA1411400, the Natural Science Foundation of China through Grant No.~12350404, the Quantum Science and Technology-National Science and Technology Major Project through Grant No.~2021ZD0302600, the Science and Technology Commission of Shanghai Municipality under Grants No.~23JC1400600 and No.~24LZ1400100, and it is sponsored by the ``Shuguang Program'' supported by the Shanghai Education Development Foundation and Shanghai Municipal Education Commission.
\end{acknowledgments}

\clearpage
\section{End matter}
\subsection{Diode efficiency factor}
We prove Eq.~(\ref{sgneta}). For clarity, we rewrite Eq.~(\ref{1st2nd}) as
\begin{equation}\label{def}
I(\phi,\delta)=S_1\sin\phi + S_2\sin(2\phi+\delta),
\end{equation}
where $S_1,S_2>0$, $\delta=\phi_2-2\phi_1$, $\phi\in[0,2\pi)$. Here, we have shifted the global phase $\phi$ to absorb the phase of the first harmonic, a transformation that leaves the critical currents $I_c^\pm$ invariant. Because the net current $I(\phi,\delta)$ vanishes when integrated $\phi$ over a full period, its absolute extremum values correspond to the forward and backward critical currents, namely $M(\delta)\equiv \max_{\phi} I(\phi,\delta)>0$ and $\min_{\phi} I(\phi,\delta)<0$. Since $I(-\phi,-\delta)=-I(\phi,\delta)$,
\begin{equation}
\min_{\phi} I(\phi,\delta)=-\max_{\phi} I(-\phi,-\delta)=-M(-\delta),
\end{equation}
so that the diode efficiency factor becomes
\begin{equation}\label{eta_M}
\eta(\delta) = \frac{M(\delta)-M(-\delta)}{M(\delta)+M(-\delta)}.
\end{equation}

For any $\delta_1$ and $\delta_2$,
\begin{equation}
\begin{aligned}
\left|I(\phi,\delta_1)-I(\phi,\delta_2)\right|&=S_2\left|\sin(2\phi+\delta_1)-\sin(2\phi+\delta_2)\right| \\
&\le S_2|\delta_1-\delta_2|,
\end{aligned}
\end{equation}
and therefore
\begin{equation}
\begin{aligned}
\left|M(\delta_1)-M(\delta_2)\right|&=\left|\max_{\phi}I(\phi,\delta_1)-\max_{\phi}I(\phi,\delta_2)\right| \\
&\le\max_{\phi}\left|I(\phi,\delta_1)-I(\phi,\delta_2)\right|\\
&\le S_2|\delta_1-\delta_2|.
\end{aligned}
\end{equation}
Thus $M(\delta)$ is globally Lipschitz continuous and hence differentiable almost everywhere.

We next locate the global maximizers $\phi_\delta\in\operatorname*{arg\,max}_{\phi}I(\phi,\delta)$. First, $\phi_\delta$ must lie in $[0,\pi]$. Indeed, if $\pi<\phi_\delta<2\pi$,
\begin{equation}
I(\phi_\delta-\pi,\delta)-I(\phi_\delta,\delta)=-2S_1\sin\phi_\delta>0,
\end{equation}
which contradicts the global maximality of $\phi_\delta$. Consider
\begin{equation}
I(\pi-\phi,\delta)-I(\phi,\delta)=-2S_2\cos\delta\,\sin(2\phi).
\label{reflection}
\end{equation}
For $\cos\delta>0$, if $\phi\in(\pi/2,\pi)$,
\begin{equation}
I(\pi-\phi,\delta)>I(\phi,\delta).
\end{equation}
Hence no global maximum can lie in $(\pi/2,\pi)$, because every such point is exceeded by its reflection in $(0,\pi/2)$. Furthermore,
\begin{equation}
\left.\frac{\partial I(\phi,\delta)}{\partial\phi}\right|_{\phi=0}=S_1+2S_2\cos\delta>0,
\end{equation}
which excludes $\phi=0$. Since $I(\pi,\delta)=I(0,\delta)$, the same local increase from $\phi=0$ also excludes $\phi=\pi$ as a global maximum. Finally,
\begin{equation}
\left.\frac{\partial I(\phi,\delta)}{\partial\phi}\right|_{\phi=\frac{\pi}{2}}=-2S_2\cos\delta\neq0,
\end{equation}
so $\phi=\pi/2$ is excluded as well. Thus $\phi_\delta\in(0,\pi/2)$. An analogous argument gives $\phi_\delta\in(\pi/2,\pi)$ for $\cos(\delta)<0$. Consequently, every global maximizer satisfies
\begin{equation}
\phi_\delta\in
\begin{cases}
(0,\pi/2), & \cos\delta>0,\\
(\pi/2,\pi), & \cos\delta<0.
\end{cases}
\label{maximizer_quadrant}
\end{equation}

At any differentiability point of $M$, Danskin's theorem implies that $\partial_\delta I$ takes the same value at all global maximizers. Thus, even when the maximizer is not unique, for any $\phi_\delta$,
\begin{equation}
M'(\delta)=\left.\frac{\partial I(\phi,\delta)}{\partial\delta}\right|_{\phi=\phi_\delta}=S_2\cos(2\phi_\delta+\delta).
\end{equation}
Since $\phi_\delta$ is an interior maximizer,
\begin{equation}
  \left.\frac{\partial I(\phi,\delta)}{\partial\phi}\right|_{\phi=\phi_\delta}=S_1\cos\phi_\delta+2S_2\cos(2\phi_\delta+\delta)=0,
\end{equation}
and hence
\begin{equation}\label{Mprime}
M'(\delta)=-\frac{S_1}{2}\cos\phi_\delta.
\end{equation}
Combining this result with Eq.~(\ref{maximizer_quadrant}) gives, almost everywhere,
\begin{equation}\label{Mprime_sign}
\operatorname{sgn}M'(\delta)=-\operatorname{sgn}(\cos\delta).
\end{equation}

At points where $M$ is differentiable at both $\delta$ and $-\delta$, differentiating Eq.~(\ref{eta_M}) gives
\begin{equation}\label{monotonicity_of_eta}
\eta'(\delta)=\frac{2\left[M'(\delta)M(-\delta)+M(\delta)M'(-\delta)\right]}{\left[M(\delta)+M(-\delta)\right]^2}.
\end{equation}
Since $M(\delta),M(-\delta)>0$ and $\operatorname{sgn}M'(-\delta)=-\operatorname{sgn}[\cos(-\delta)]= -\operatorname{sgn}(\cos\delta)$, it follows that, almost everywhere,
\begin{equation}\label{monotonicity}
\operatorname{sgn}\eta'(\delta)=-\operatorname{sgn}(\cos\delta).
\end{equation}
Because the denominator in Eq.~(\ref{eta_M}) has a positive uniform lower bound, $\eta(\delta)$ is also globally Lipschitz continuous and therefore absolutely continuous. Together with the almost-everywhere sign relation above, this implies that $\eta$ is strictly decreasing on each interval with $\cos\delta>0$ and strictly increasing on each interval with $\cos\delta<0$. Finally, $\eta(n\pi)=0$ for $n\in\mathbb{Z}$, and the above monotonicity yields
\begin{equation}
\operatorname{sgn}\eta(\delta)=-\operatorname{sgn}(\sin\delta),
\end{equation}
which proves Eq.~(\ref{sgneta}). Moreover, Eq.~(\ref{monotonicity}) captures the monotonic evolution of $\eta$ with $\delta$ and the resulting maxima of $|\eta|$ at $\delta=\frac{\pi}{2}~(\mathrm{modulo}~\pi)$, in full agreement with the numerical results in Fig.~\ref{fig2}(a).

\subsection{The harmonic-selection rule from local bound-state}
Here we derive the harmonic-selection rule analytically in the local-bound-state limit. For simplicity, we assume a positive vortex number $p>0$. Squaring the continuum Andreev Hamiltonian in Eq.~(\ref{ana_H}) yields the eigenvalue problem 
\begin{equation}\label{square}
  \left[-v^2 \partial_\theta^2 + m^2(\theta,\phi)\mp v m'(\theta,\phi)\right] \chi_\pm = E^2(\phi)\chi_\pm,
\end{equation}
where $m(\theta,\phi)=\Delta(\theta)\cos\left(\frac{\phi+p\theta}{2}\right)$, $m'(\theta,\phi)=\partial_\theta m(\theta,\phi)$, $\chi_\pm=u_1\pm u_2$, and $\Psi=(u_1,u_2)^T$. The nodes of the position-dependent Andreev coupling $m(\theta,\phi)$ correspond to the centers of the Josephson vortices, located at
\begin{equation}\label{theta_i}
  \theta_i(\phi)=\frac{\pi+2\pi i-\phi}{p}, \qquad i=0,1,\ldots,p-1.
\end{equation}
Expanding the mass profile near the $i$-th vortex center ($\theta=\theta_i+\zeta$) yields the linear approximation $m(\theta,\phi)\sim\lambda_i(\phi)\zeta$, where $\lambda_i(\phi)=-\frac{p}{2}(-1)^i\Delta(\theta_i)$~\cite{volovik1999fermion,grosfeld2011observing,potter2013anomalous}. Substituting this linearized mass profile into Eq.~(\ref{square}) reduces the system to a shifted quantum harmonic oscillator:
\begin{equation}\label{oscillator}
  \left[-v^2\partial_\zeta^2+\lambda_i^2\zeta^2\mp v\lambda_i\right]\chi_\pm =E^2\chi_\pm .
\end{equation}
The resulting localized discrete energy levels are solved analytically as
\begin{equation}\label{spectrum}
  \varepsilon_{i,n}(\phi)=\pm\sqrt{2nv|\lambda_i(\phi)|},  \qquad n=0,1,2,\ldots.
\end{equation}
Because the total Josephson current Eq.~(\ref{current}) is evaluated as a sum over positive Andreev levels, and the strict localized limit yields a dispersionless zero mode, we focus exclusively on the positive-energy branch ($n\geq1$):
\begin{equation}\label{positive}
  \varepsilon^+_{i,n}(\phi)=\sqrt{nvp\,\Delta(\theta_i)},\qquad n\geq1 .
\end{equation}
In the weak-inhomogeneity limit ($\Delta_j\ll\Delta_0$), we utilize the Fourier expansion of the pairing profile Eq.~(\ref{Delta}) to expand Eq.~(\ref{positive}) as
\begin{equation}\label{eq:expansion}
  \varepsilon^+_{i,n}(\phi)\sim\varepsilon_n^{(0)}\left[1+\frac{1}{2\Delta_0}\sum_{j>0}\Delta_j\cos(j\theta_i+\varphi_j)\right],
\end{equation}
where $\varepsilon_n^{(0)}=\sqrt{nvp\Delta_0}$. For each index $n$, there are $p$ nearly degenerate positive-energy states distributed across the vortex centers. This expression directly elucidates the low-energy spectral modulations observed in Figs.~\ref{fig3}(e) and~\ref{fig3}(f). As the global phase difference $\phi$ is varied, the vortex centers $\theta_i(\phi)$ adiabatically translate around the annulus, causing each localized level to sample the underlying pairing landscape. Consequently, for the two-site perturbation profile considered in our edge model, the low-energy branches undergo pronounced modulations whenever a vortex center passes through a localized pairing perturbation. The summed energy of this local multiplet is obtained by summing over all vortex positions:
\begin{equation}\label{sumenergy}
  E_n^{\mathrm{loc}}(\phi)=\sum_{i=0}^{p-1}\varepsilon^+_{i,n}(\phi) \sim p\varepsilon_n^{(0)}+\frac{\varepsilon_n^{(0)}}{2\Delta_0}\sum_{j>0}\Delta_j\mathcal S_j(\phi),
\end{equation}
where $\mathcal S_j(\phi)=\sum_{i=0}^{p-1}\cos\left[j\theta_i(\phi)+\varphi_j\right]$. Using Eq.~(\ref{theta_i}), the sum evaluates to
\begin{equation}\label{selection}
  \mathcal S_j(\phi)=
  \begin{cases}
  p\cos\left(\ell\pi-\ell\phi+\varphi_{\ell p}\right), & j=\ell p,\\
  0, & j\notin p\mathbb Z .
  \end{cases}
\end{equation}
Thus only spatial harmonics $j=\ell p$ survive after summation over all vortex positions. Substituting into Eq.~(\ref{sumenergy}) gives
\begin{equation}\label{local_energy}
  E_n^{\mathrm{loc}}(\phi) \sim p\varepsilon_n^{(0)}+ \frac{p\varepsilon_n^{(0)}}{2\Delta_0} \sum_{\ell>0}\Delta_{\ell p} \cos\left(\ell\phi+\tilde{\phi}_{\ell p}\right),
\end{equation}
where $\tilde{\phi}_{\ell p}=-\varphi_{\ell p}-\ell\pi$. Eq.~(\ref{local_energy}) constitutes the localized-bound-state manifestation of our harmonic-selection rule: a structural spatial harmonic with $j=\ell p$ explicitly generates the $\ell$-th phase harmonic of the Josephson energy. The resulting contribution to the Josephson current is evaluated via Eq.~(\ref{current}):
\begin{equation}\label{local_current}
  I_n^{\mathrm{loc}}(\phi)\sim\frac{e p\varepsilon_n^{(0)}}{\hbar\Delta_0}  \sum_{\ell>0}\ell\Delta_{\ell p}\sin\left(\ell\phi+\tilde{\phi}_{\ell p}\right).
\end{equation}
Thus, the selected spatial harmonic $j=\ell p$ produces the $\ell$th harmonic of the CPR.

\end{document}